\documentclass[baaa]{baaa}

\usepackage[pdftex]{hyperref}
\usepackage{subfigure}
\usepackage{natbib}
\usepackage{helvet,soul}
\usepackage[font=small]{caption}

\begin{document}


\journalvol{60}
\journalyear{2018}
\journaleditors{P. Benaglia, A.C. Rovero, R. Gamen \& M. Lares}


\contriblanguage{0}


\contribtype{1}

\thematicarea{1}

\title{Caracterización de galaxias elípticas\\ en contexto cosmológico}
\subtitle{}

\titlerunning{Galaxias elípticas en contexto cosmológico}


\author{L.J. Zenocratti\inst{1}, M.E. De Rossi\inst{2,3}, A.V. Smith Castelli\inst{4,5}, F.R. Faifer\inst{1,5} }
\authorrunning{Zenocratti et al.}


\contact{lucashawk6@hotmail.com}

\institute{Facultad de Ciencias Astron\'omicas y Geof\'isicas, UNLP, Paseo del Bosque s/n, B1900FWA, La Plata, Argentina \and
Universidad de Buenos Aires, Facultad de Ciencias Exactas y Naturales y Ciclo B\'asico Com\'un. Buenos Aires, Argentina  \and
CONICET-Universidad de Buenos Aires, Instituto de Astronom\'{\i}a y F\'{\i}sica del Espacio (IAFE). Buenos Aires, Argentina \and
Consejo Nacional de Investigaciones Cient\'ificas y T\'ecnicas, Argentina \and
Instituto de Astrof\'isica de La Plata, UNLP, CONICET, Paseo del Bosque s/n, B1900FWA Argentina}


\resumen{
Presentamos resultados preliminares de un proyecto destinado a explorar
galaxias elípticas en simulaciones numéricas en contexto cosmológico,
con el fin de proveer posibles escenarios de formación para muestras de 
galaxias observadas.
Para ello, realizamos un estudio de las propiedades integradas de tales 
sistemas simulados en función de su masa y a corrimiento al rojo $z=0$. Se encontraron algunas diferencias entre el diagrama color-magnitud simulado y resultados observacionales del cúmulo de Virgo.}

\abstract{
We present preliminary results from a project aimed at exploring elliptical galaxies in numerical simulations within cosmological context to provide possible formation scenarios for samples of observed galaxies.
We studied the integrated properties of such simulated systems in terms of mass and at redshift $z=0$. Some differences were found between the simulated color-magnitude diagram and observational results in the Virgo cluster.
}



\maketitle
\section{Introducción}
\label{S_intro}

Las galaxias elípticas constituyen la población más numerosa que es 
posible encontrar en cúmulos y grupos en el Universo Local, y su estudio puede aportar claves únicas para entender el proceso de formación de estructuras en el Universo. Sin embargo, pese a los esfuerzos realizados en este sentido, aún se carece de un escenario que pueda
explicar consistentemente todas las propiedades observadas de tales sistemas en forma global en los escenarios de formación de galaxias \citep{kormendy_galaxies_2009,silchenko_galaxies_2012}. Por ejemplo, un tema actualmente en estudio es la existencia de dicotomías dentro de la familia de las galaxias elípticas, como ser, entre rotadores rápidos y lentos, y galaxias elípticas normales y enanas \citep{cappellari_earlytype_2016,schombert_dichotomy_2017}.

En este trabajo, mostramos resultados preliminares de un estudio estadístico de las propiedades de galaxias elípticas simuladas, focalizándonos en el diagrama color-magnitud de dichos sistemas a corrimiento al rojo $z=0$, y su comparación con resultados observacionales. El objetivo específico de este estudio es analizar las propiedades de la componente estelar de galaxias mediante simulaciones y compararlas con observaciones. Esto permitirá establecer escenarios de formación y evolución para las galaxias elípticas, y entender los procesos que dan lugar a las propiedades observadas de estas, teniendo en cuenta el medio en el que habitan.

\section{Descripción de la simulación utilizada}
Se utilizaron los catálogos de galaxias generados mediante la aplicación del modelo semi-analítico de \citet{henriques_model_2015} sobre la simulación del Millennium\footnote[1]{\url{http://galformod.mpa-garching.mpg.de/portal/}} \citep{springel_millennium_2005}. Este modelo incluye prescripciones físicas para procesos como enfriamiento del gas, formación estelar, retroalimentación al medio por supernovas y núcleos activos de galaxias, e interacciones y fusiones entre sistemas galácticos, entre otros.

La simulación empleada fue obtenida adaptando la versión original de la simulación del Millennium a los primeros datos de la Colaboración Planck (2014). La cosmología de Planck adopta los siguientes parámetros cosmológicos: $\Omega_{\Lambda}=0.685$, $\Omega_\textrm{m}=0.315$, $\Omega_\textrm{b}=0.049$ y $h=0.673$. La simulación del Millennium traza $2\,160^3$ partículas desde $z=127$ hasta el presente, siguiendo la evolución de halos de materia oscura cuyas galaxias asociadas abarcan hasta 5 órdenes de magnitud en masa estelar para $z=0$. El volumen simulado corresponde a una caja cúbica de aproximadamente 714~Mpc de lado, con una resolución en masa de $1.43\times10^{9}$ M$_\odot$. 
 
\section{Selección de galaxias}

En este trabajo, nos enfocamos en el análisis de galaxias elípticas, obtenidas del modelo semi-analítico mencionado en la sección anterior. Para esto, estudiamos las galaxias que tienen una masa estelar $\textrm{M}_*\geq10^9~\textrm{M}_\odot$, corte que permite seleccionar sistemas con más de
125 partículas de materia oscura, en promedio.  Utilizando datos provenientes de la simulación
del Millennium-II \citep{boylan_2009}, la cual presenta mayor resolución pero menor volumen, se obtuvo convergencia numérica para nuestros resultados dentro del rango de masa estelar seleccionado. 
Para mayor claridad, en este trabajo no se incluyen resultados de esta última simulación.  
Para asegurarnos de estudiar galaxias dominadas por el bulbo, 
se restringió la muestra a aquellos sistemas simulados que presentan más del 90~\% de su masa estelar en el bulbo. Por último, si bien nuestro estudio pretende analizar galaxias simuladas a distintos corrimientos al rojo, para esta presentación nos focalizamos en los resultados obtenidos a $z=0$.

\begin{figure}[!t]
  \centering
  \includegraphics[width=0.4\textwidth]{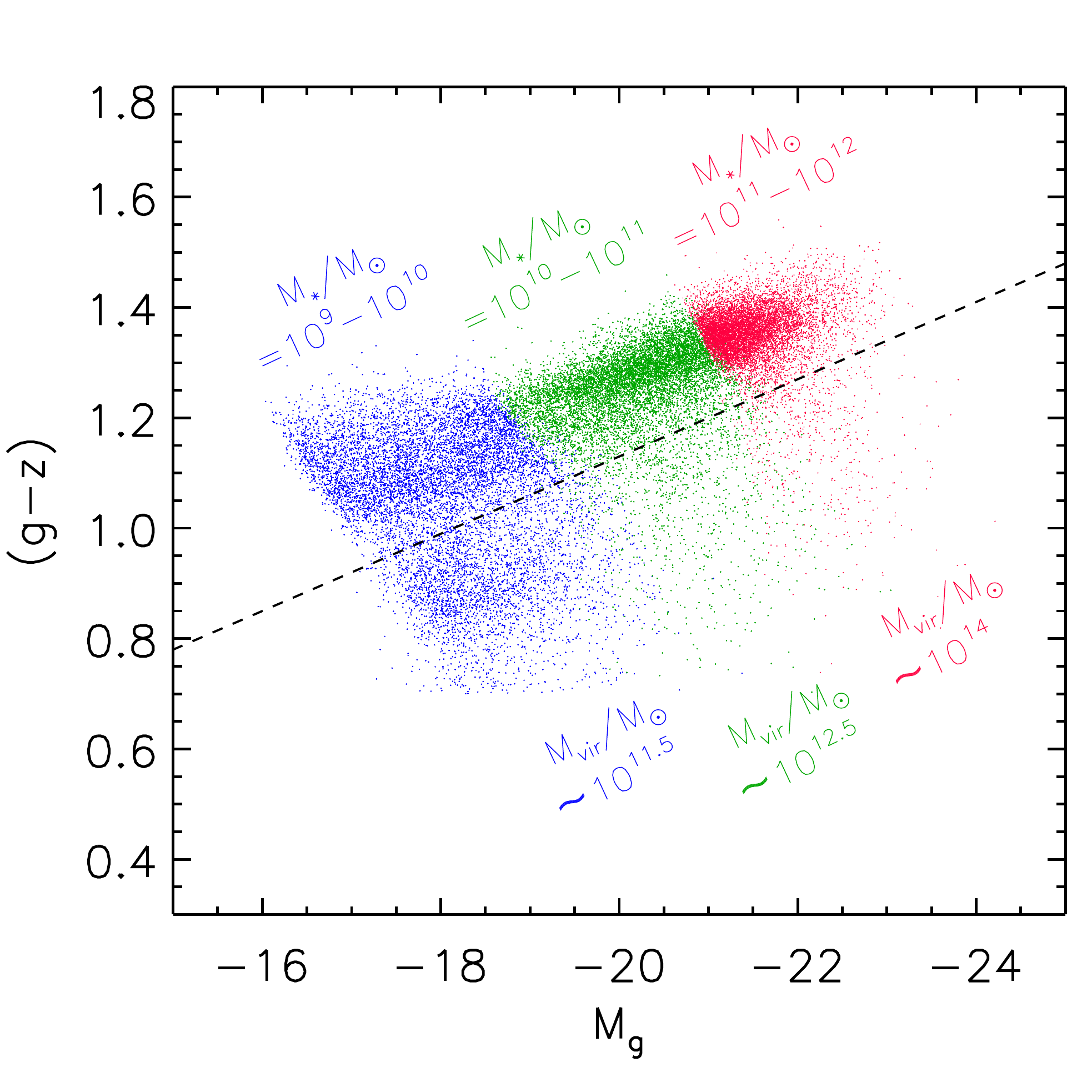}
  \caption{Diagrama color-magnitud para las galaxias simuladas seleccionadas en tres rangos de masa distintos. La línea de trazos (negra) separa el diagrama en dos regiones, denominadas ``banda'' (por encima de la recta) y ``nube difusa'' (por debajo de la recta).
}
  \label{F_1}
\end{figure}

\section{Resultados}
Como se mencionó anteriormente, nuestro objetivo es comparar las simulaciones con las observaciones. Con este fin, comenzamos el estudio a partir del diagrama color-magnitud, ya que dicho diagrama es una de las herramientas observacionales más importantes con que se cuenta. Analizamos inicialmente galaxias a $z=0$ para su posterior comparación con observaciones. En la Figura~\ref{F_1}, se muestra el diagrama color-magnitud para la muestra de galaxias seleccionadas de la simulación. Las cantidades usadas incluyen efectos de extinción por polvo. Las magnitudes corresponden a los filtros g ($\lambda=0.469~{\mu}\textrm{m}$) y z ($\lambda=0.893~{\mu}\textrm{m}$) del {\em Sloan Digital Sky Survey} (SDSS, por sus siglas en inglés), respectivamente, definido por \citet{fukugita_filters_1996}. El modelo de síntesis poblacional aplicado sobre los catálogos de galaxias es el dado por \citet{maraston_evolution_2005}, con una función inicial de masa (IMF, por sus siglas en inglés) de \citet{chabrier_imf_2003}.

Como es posible apreciar en la Figura~\ref{F_1}, la muestra fue separada en tres rangos de masa estelar, y se puede observar que cada rango ocupa una región específica del diagrama: a medida que la masa estelar aumenta, los colores se enrojecen y los brillos aumentan. El enrojecimiento se debe principalmente a un aumento en la metalicidad y posiblemente, en segundo orden, a efectos de edad. En la mencionada figura, se indica además la masa virial (masa del halo de materia oscura, M$_\textrm{vir}$) asociada a cada rango de masa estelar. Se ve que la masa estelar de las galaxias está muy vinculada a la masa de sus halos de materia oscura. 

En la Figura~\ref{F_1}, queda en evidencia que la gran mayoría de las galaxias seleccionadas se encuentra en una ``banda'' bien poblada, y que por debajo de esta hay una ``nube difusa'' de galaxias. Para establecer qué tipo de galaxias hay en cada una de estas dos regiones, las separamos gráficamente mediante la recta de ecuación $\textrm{(g-z)}=-0.07\textrm{M}_\textrm{g}-0.3$ (línea de trazos negra en la figura), y analizamos dos características fundamentales de las galaxias elípticas: la fracción de gas frío y la tasa de formación estelar ({\em Star Formation Rate}, o SFR por sus siglas en inglés). 

\begin{figure}[!t]
  \centering
  \includegraphics[width=0.4\textwidth]{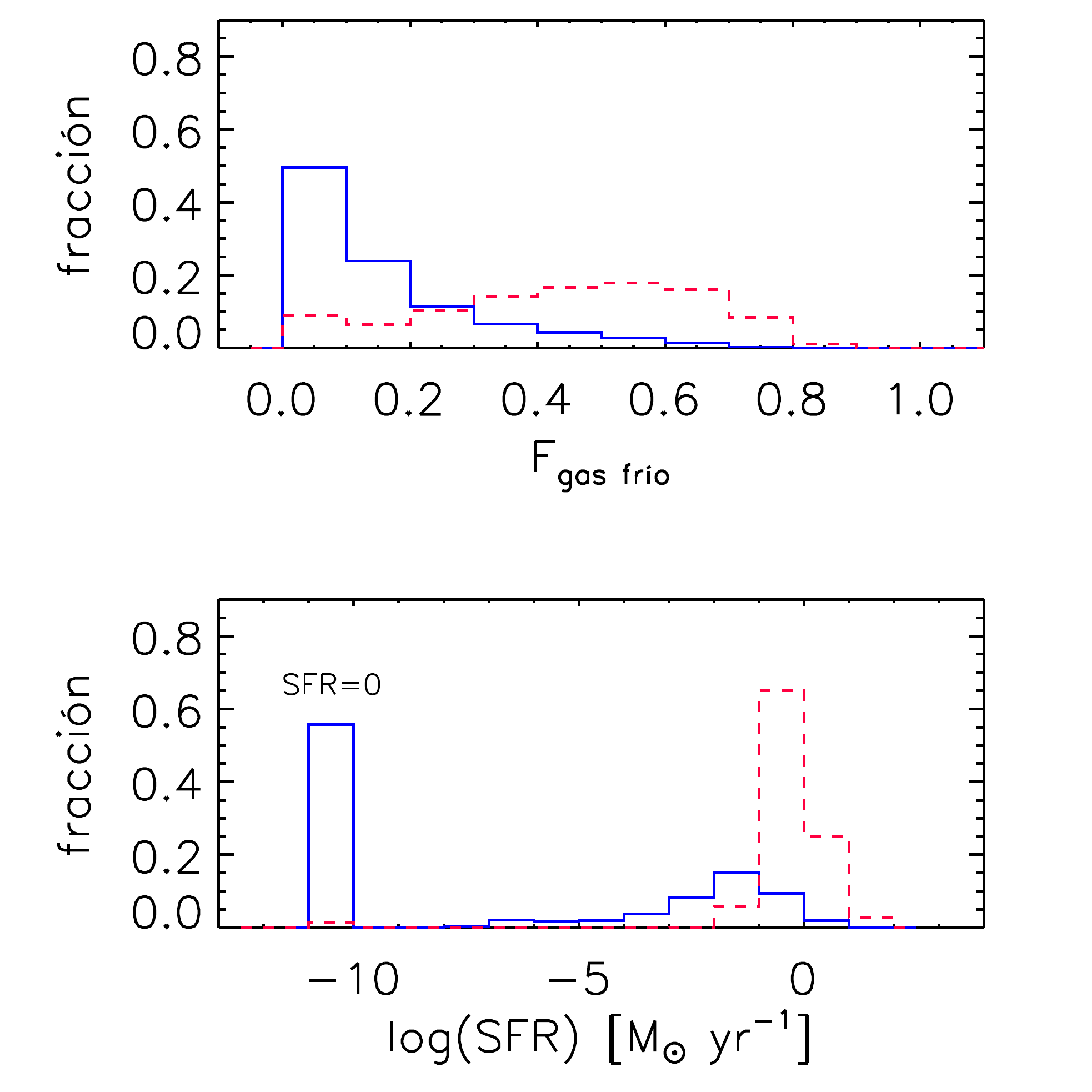}
  \caption{Histogramas de fracción de gas frío (panel superior) y de tasa de formación estelar (panel inferior) para galaxias simuladas. La línea sólida (azul) corresponde a la ``banda'' bien poblada del diagrama color-magnitud, y posiblemente se trate en su mayoría de galaxias elípticas; la línea de trazos (rojo) corresponde a la \``nube difusa''.
}
  \label{F_2}
\end{figure}

Las galaxias elípticas se caracterizan por presentar bajos porcentajes de gas frío, y además muy bajas tasas de formación estelar. La Figura~\ref{F_2} muestra histogramas de fracción de gas frío (panel superior) y de tasa de formación estelar (panel inferior), separando las galaxias de la ``banda'' bien poblada (línea sólida, en azul) y de la ``nube difusa'' (línea de trazos, en rojo). La fracción de gas frío se calcula respecto de la masa total de gas frío y masa total de estrellas. Se puede ver que la mayoría de las galaxias de la primer región cumplen las dos condiciones mencionadas para galaxias elípticas. Así entonces, se puede decir que estas galaxias son, en principio, galaxias elípticas. 

\begin{figure}[!b]
  \vspace{-10mm}
  \centering
  \includegraphics[width=0.45\textwidth]{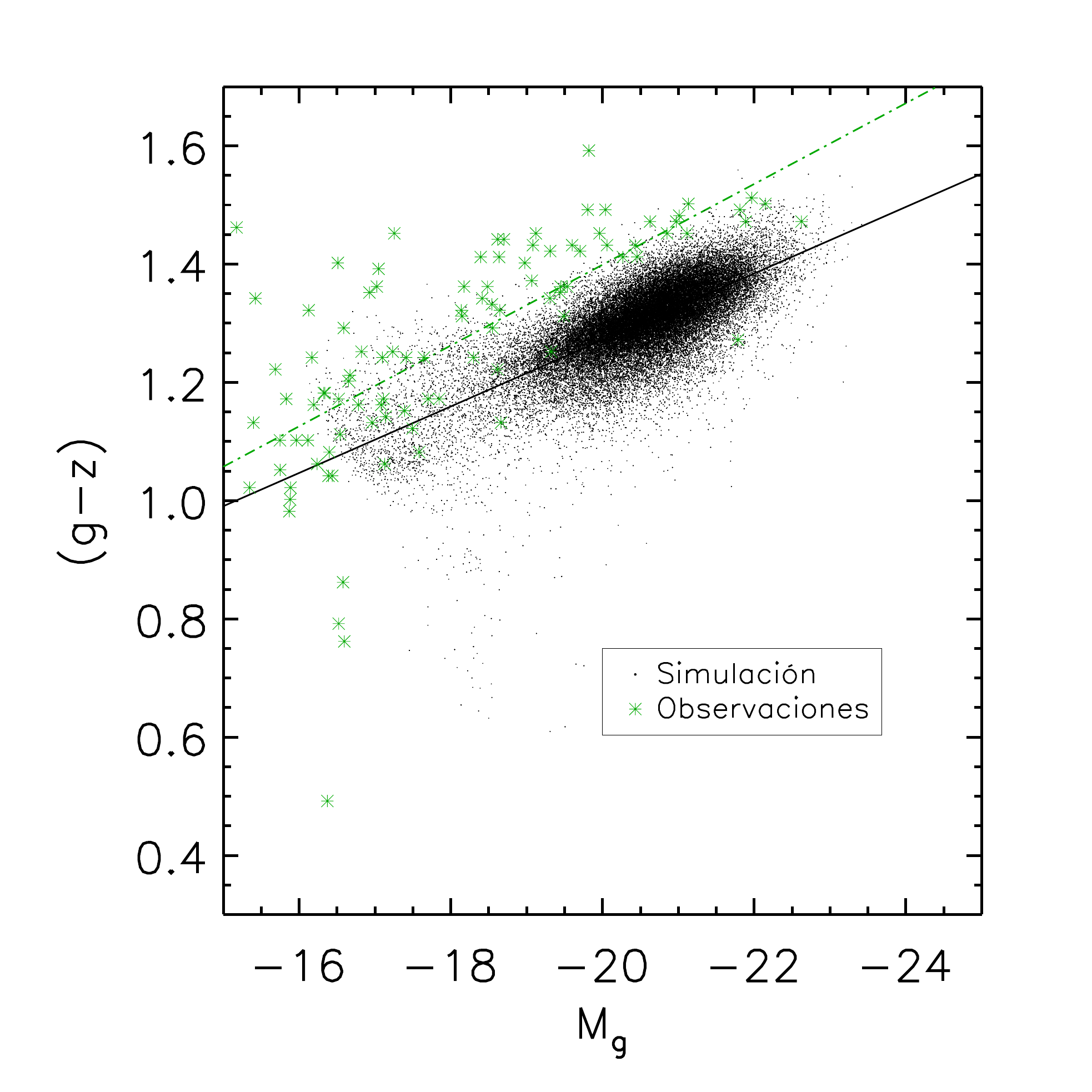}
  \vspace{-4mm}
  \caption{Diagrama color-magnitud comparando la simulación con las observaciones. La línea sólida (negra) representa un ajuste lineal a la relación color-magnitud para las galaxias simuladas; la línea de trazo y punto (verde) corresponde a la relación para galaxias observadas del cúmulo de Virgo \citep{chen_2010}.
}
  \label{F_3}
\end{figure}

Teniendo en cuenta los histogramas mencionados, confeccionamos un nuevo diagrama color-magnitud donde se incluyen únicamente las galaxias que cumplen con F$_\textrm{gas frío}<0.1$ y $\textrm{SFR}<1~\textrm{M}_\odot~\textrm{yr}^{-1}$. De la Figura \ref{F_2}, se ve que no hay un porcentaje significativo de galaxias con alta tasa de formación estelar; a pesar de esto, el corte en SFR se aplicó porque permite descartar objetos cuyos colores integrados resultan significativamente más azules que la secuencia roja, debido a la formación estelar. En este nuevo diagrama, las galaxias seleccionadas ocupan una zona bien definida (ver Figura~\ref{F_3}), correspondiendo la muestra a aproximadamente $150\,000$ galaxias. Esto indica que la utilización de los criterios antes mencionados nos proporciona una muestra muy posiblemente dominada por galaxias elípticas. Un paso a futuro incluirá la utilización y análisis exhaustivo de propiedades integradas de las poblaciones estelares de la muestra seleccionada, tales como la edad y la metalicidad, para caracterizar qué tipo de sistemas galácticos la componen.

Con las galaxias de baja fracción de gas frío y de baja tasa de formación estelar que hemos seleccionado, realizamos una comparación con datos observacionales de galaxias del cúmulo de Virgo del trabajo de \citet{chen_2010}. La Figura~\ref{F_3} muestra el diagrama color-magnitud de las galaxias simuladas junto con las galaxias observadas; la relación color-magnitud se ajustó en ambos casos con una función lineal mediante mínimos cuadrados. La línea sólida muestra el ajuste para la simulación, mientras que la línea de trazo y punto corresponde a las observaciones. En esta última figura, se ve que la \textrm{relación color-magnitud} simulada no coincide exactamente con las observaciones: para la simulación, la pendiente del ajuste lineal es de $-0.06\pm0.01$, mayor que la correspondiente a las observaciones, que vale $-0.08\pm0.01$. Las ordenadas al origen también son distintas para la simulación y las observaciones, siendo estas $0.15\pm0.003$ y $0.03\pm0.14$, respectivamente. Además, la mayoría de las galaxias simuladas parecen ser más brillantes en el filtro g y menos rojas que las observadas. Es importante mencionar que la comparación entre simulación y observaciones se realiza solo en el intervalo de luminosidades donde hay superposición en las muestras. Comparaciones más exhaustivas se llevarán a cabo en un futuro trabajo.

\section{Conclusiones y trabajo a futuro}

De los resultados obtenidos, se puede concluir que las galaxias seleccionadas de la simulación del Millennium muestran características fotométricas distintas a las galaxias elípticas que definen la relación color-magnitud observada. Estas diferencias pueden deberse a discrepancias en los parámetros fotométricos de referencia utilizados tanto para la simulación como para las observaciones, o bien a las correcciones que se aplican en las magnitudes. 

Seguiremos estudiando los modelos utilizados en las simulaciones para determinar las principales prescripciones físicas que pudieron afectar a la muestra de galaxias simuladas, y poder así determinar a qué se deben las diferencias mencionadas. De esta forma, esperamos poder refinar la muestra y delimitar con mayor precision las características fundamentales de las galaxias simuladas, lo cual nos permitirá realizar una comparación más exhaustiva con muestras observacionales. Una vez cumplimentado este objetivo, trabajaremos con muestras simuladas a distintos corrimientos al rojo para poder analizar la evolución de estos sistemas en función del ambiente en el que habitan.  

\label{thanks}
\begin{acknowledgement}
Agradecemos a la Asociación Argentina de Astronomía por el espacio cedido para mostrar nuestros resultados. Agradecemos a los subsidios PICT-2015-3125 de la ANPCyT, PIP 112-201501-00447 del CONICET, y UNLP G151 de la UNLP (Argentina). Las bases de datos de la simulación del Millennium usadas en esta publicación y la aplicación web que provee acceso a estas fueron construidas como parte de las actividades del {\em German Astrophysical Virtual Observatory (GAVO)}. 

\end{acknowledgement}


\bibliographystyle{baaa}
\small
\bibliography{biblio}

\begin{thebibliography}{}
\catcode`\~=\active \def~{\penalty10000 \ }

\bibitem[\protect\citeauthoryear{{Boylan-Kolchin}, {Springel}, {White},
  {Jenkins} \& {Lemson}}{{Boylan-Kolchin} et al.}{2009}]{boylan_2009}
{Boylan-Kolchin} M.,  et al., 2009, \mnras, 398, 1150

\bibitem[\protect\citeauthoryear{{Cappellari}}{{Cappellari}}{2016}]{cappellari_earlytype_2016}
{Cappellari} M.,  2016, \araa, 54, 597

\bibitem[\protect\citeauthoryear{{Chabrier}}{{Chabrier}}{2003}]{chabrier_imf_2003}
{Chabrier} G.,  2003, \pasp, 115, 763

\bibitem[\protect\citeauthoryear{{Chen}, {C{\^o}t{\'e}}, {West}, {Peng} \&
  {Ferrarese}}{{Chen} et al.}{2010}]{chen_2010}
{Chen} C.-W.,  et al., 2010, \apjs, 191, 1

\bibitem[\protect\citeauthoryear{{Fukugita}, {Ichikawa}, {Gunn}, {Doi},
  {Shimasaku} \& {Schneider}}{{Fukugita} et al.}{1996}]{fukugita_filters_1996}
{Fukugita} M.,  et al., 1996, \aj, 111, 1748

\bibitem[\protect\citeauthoryear{{Henriques}, {White}, {Thomas}, {Angulo},
  {Guo}, {Lemson}, {Springel} \& {Overzier}}{{Henriques} et
  al.}{2015}]{henriques_model_2015}
{Henriques} B.~M.~B.,  et al., 2015, \mnras, 451, 2663

\bibitem[\protect\citeauthoryear{{Kormendy}, {Fisher}, {Cornell} \&
  {Bender}}{{Kormendy} et al.}{2009}]{kormendy_galaxies_2009}
{Kormendy} J.,  et al., 2009, \apjs, 182, 216

\bibitem[\protect\citeauthoryear{{Maraston}}{{Maraston}}{2005}]{maraston_evolution_2005}
{Maraston} C.,  2005, \mnras, 362, 799

\bibitem[\protect\citeauthoryear{{Schombert}}{{Schombert}}{2016}]{schombert_dichotomy_2017}
{Schombert} J.~M.,  2016, \aj, 152, 214

\bibitem[\protect\citeauthoryear{{Sil'chenko}, {Proshina}, {Shulga} \&
  {Koposov}}{{Sil'chenko} et al.}{2012}]{silchenko_galaxies_2012}
{Sil'chenko} O.~K.,  et al., 2012, \mnras, 427, 790

\bibitem[\protect\citeauthoryear{{Springel}, {White}, {Jenkins}, {Frenk},
  {Yoshida}, {Gao}, {Navarro}, {Thacker}, {Croton}, {Helly}, {Peacock}, {Cole},
  {Thomas}, {Couchman}, {Evrard}, {Colberg} \& {Pearce}}{{Springel} et
  al.}{2005}]{springel_millennium_2005}
{Springel} V.,  et al., 2005, \nat, 435, 629

\end{thebibliography}


\end{document}